\documentclass[10pt,conference]{IEEEtran}
\IEEEoverridecommandlockouts{}
\usepackage{cite}
\usepackage{amsmath,amssymb,amsfonts}
\usepackage{algorithmic}
\usepackage{graphicx}
\usepackage{textcomp}
\usepackage{xcolor}
\usepackage{float}
\usepackage{subfig}
\usepackage{dsfont}
\usepackage{breqn}
\usepackage[draft]{hyperref}
\usepackage{verbatim}
\usepackage{mathrsfs}

\usepackage{booktabs}

\usepackage{xcolor}

\newcommand{\systemname}{ReinWiFi}

\DeclareMathOperator*{\argmin}{arg\,min}

\newcommand{\BE}{\mathtt{BE}}

\newcommand{\VI}{\mathtt{VI}}

\newcommand{\devSet}{\mathcal{U}}
\newcommand{\devNum}{U}
\newcommand{\linkset}{\mathcal{L}}
\newcommand{\taskset}{\mathcal{T}}
\newcommand{\thruTS}{\taskset_{\link}^{f}}
\newcommand{\rttTS}{\taskset_{\link}^{r}}

\newcommand{\m}{m}
\newcommand{\n}{n}
\newcommand{\devi}{i}
\newcommand{\devj}{j}
\newcommand{\dataSetTau}{\tau}
\newcommand{\dev}{u}
\newcommand{\slotlen}{T_s}
\newcommand{\link}{{\devi},{\devj}}
\newcommand{\linkIdx}{({\devi},{\devj})}
\newcommand{\thru}{r}
\newcommand{\thruLimit}{b}
\newcommand{\rtt}{d}
\newcommand{\rttLimit}{\mathrm{D}}
\newcommand{\avgRTT}{\bar{\rtt}}
\newcommand{\avgThru}{\bar{\thru}}
\newcommand{\varRTT}{\sigma_{\rtt}}
\newcommand{\varThru}{\sigma_{\thru}}
\newcommand{\eleVal}[2]{#1_{\link}^#2}
\newcommand{\eleRTT}{\eleVal{\rtt}{{\m}}}
\newcommand{\eleRTTClassify}{\eleVal{\rtt}{{\m, \classifyFlag}}}
\newcommand{\eleRTTAlt}{\eleVal{\rtt}{{\n}}}
\newcommand{\eleThru}{\eleVal{\thru}{{\m}}}
\newcommand{\eleThruClassify}{\eleVal{\thru}{{\m, \classifyFlag}}}
\newcommand{\elePreRTT}{\eleVal{\hat{\rtt}}{{\n}}}
\newcommand{\elePreThru}{\eleVal{\hat{\thru}}{{{\m}}}}
\newcommand{\eleThruLimit}{\eleVal{\thruLimit}{{\m}}}
\newcommand{\eleRTTLimit}{\eleVal{\rttLimit}{{\m}}}

\newcommand{\observation}{\mathcal{O}}
\newcommand{\action}{\mathcal{A}}
\newcommand{\actionSpace}{\mathscr{A}}
\newcommand{\state}{\mathcal{S}}
\newcommand{\extendState}{\hat{\state}}

\newcommand{\policy}{\Omega}
\newcommand{\cost}{c}
\newcommand{\averageCost}{\bar{C}}

\newcommand{\stateSize}{N}
\newcommand{\weight}{\omega}

\newcommand{\trainingDataSet}{\mathscr{S}^{\trainFlag}}
\newcommand{\classifyDataSet}{\mathscr{S}^{\classifyFlag}}
\newcommand{\trainingDataSetwithRawPerInd}{\Tilde{\mathscr{S}}^{\trainFlag}}
\newcommand{\dataSetSize}{M}
\newcommand{\classifyFlag}{p}
\newcommand{\trainFlag}{s}
\newcommand{\classifyIdx}{k}
\newcommand{\classifyNum}{K}
\newcommand{\classifyCenter}{\mu}
\newcommand{\vectorization}{\phi}
\newcommand{\performanceRegion}{\psi}
\newcommand{\rawPerRegion}{\hat{\performanceRegion}}

\newcommand{\classifyAction}{\action^{\classifyFlag}}
\newcommand{\classifyObs}{\observation^{\classifyFlag}}
\newcommand{\trainObs}{\observation^{\trainFlag}}
\newcommand{\trainAction}{\action^{\trainFlag}}

\newcommand{\iterOver}[2]{#1 \in{#2}}
\newcommand{\iterLink}{\iterOver{\linkIdx}{\linkset}}
\newcommand{\iterTaskRTT}{\iterOver{{\m}}{\rttTS}}
\newcommand{\iterTaskRTTAlt}{\iterOver{{\n}}{\rttTS}}
\newcommand{\iterTaskThru}{\iterOver{{\m}}{\thruTS}}
\newcommand{\iterFromTo}[2]{#1, \ldots, #2}
\newcommand{\iterCC}{\iterFromTo{\classifyCenter_1}{\classifyCenter_{\classifyNum}}}
\newcommand{\iterCCOpt}{\iterFromTo{\classifyCenter_1^*}{\classifyCenter_{\classifyNum}^*}}
\newcommand{\iterFromPassTo}[3]{#1, #2, \ldots, #3}
\newcommand{\iterDataSet}{\iterFromPassTo{1}{2}{\dataSetSize}}
\newcommand{\iterDeviceSet}{\iterFromPassTo{0}{1}{\devNum-1}}

\newcommand{\iterCCIdx}{\iterFromPassTo{1}{2}{\classifyNum}}

\newcommand{\imitatorLoss}{ L^I }
\newcommand{\ctlLoss}{ L^q }
\newcommand{\uniformDis}{ \mathrm{Unif} }
\newcommand{\netParam}{ \boldsymbol{\theta} }
\newcommand{\inetParam}{ \netParam^w }
\newcommand{\qnetParam}{ \netParam^q }
\newcommand{\qtnetParam}{ \netParam^{q,-}}

\newcommand{\firExpLink}{ u_1,u_0 }
\newcommand{\secExpLink}{ u_2,u_0 }
\newcommand{\thiExpLink}{ u_3,u_0 }
\newcommand{\arrival}{\lambda}
\newcommand{\counter}{i}

\usepackage{titlesec}
\titlespacing\section{0pt}{10pt plus 4pt minus 2pt}{0pt plus 2pt minus 2pt}
\titlespacing\subsection{0pt}{9pt plus 4pt minus 2pt}{0pt plus 2pt minus 2pt}
\titlespacing\subsubsection{0pt}{9pt plus 4pt minus 2pt}{0pt plus 2pt minus 2pt}
\newtheorem{Definition}{Definition}

\newtheorem{Remark}{Remark}

\def\BibTeX{{\rm B\kern-.05em{\sc i\kern-.025em b}\kern-.08em
    T\kern-.1667em\lower.7ex\hbox{E}\kern-.125emX}}
\begin{document}

\title{{\systemname}: Application-Layer QoS Optimization of WiFi Networks with Reinforcement Learning\\
}
\author{
    Qianren Li, Bojie Lv, Yuncong Hong, and Rui Wang 
    \\ {Southern University of Science and Technology}
	\thanks{
	    Rui Wang is the corresponding author. \par 
        This work is supported in part by National Natural Science Foundation of China (NSFC) under Grant 62171213, in
        part by Shenzhen Science and Technology Program under Grant JCYJ20241202125328038, and in part by High Level of Special Funds under Grant G03034K004.
	}
}

\maketitle
\begin{abstract}
    The enhanced distributed channel access (EDCA) mechanism is used in current wireless fidelity (WiFi) networks to support priority requirements of heterogeneous applications. However, the EDCA mechanism can not adapt to particular quality-of-service (QoS) objective, network topology, and interference level. 
    In this paper, a novel reinforcement-learning-based scheduling framework is proposed and implemented to optimize the application-layer quality-of-service (QoS) of a WiFi network with commercial adapters and unknown interference. 
    Particularly, application-layer tasks of file delivery and delay-sensitive communication are jointly scheduled by adjusting the contention window sizes and application-layer throughput limitation, such that the throughput of the former and the round trip time of the latter can be optimized. Due to the unknown interference and vendor-dependent implementation of the WiFi adapters, the relation between the scheduling policy and the system QoS is unknown. Hence, a reinforcement learning method is proposed, in which a novel Q-network is trained to map from the historical scheduling parameters and QoS observations to the current scheduling action. It is demonstrated on a testbed that the proposed framework can achieve a significantly better performance than the EDCA mechanism.
\end{abstract}

\section{Introduction}

Reinforcement learning (RL) for radio resource management has been attracting tremendous attention since it is a promising technique to tackle unknown system statistics and solve the prohibitive policy optimization problem with tolerable complexity and good performance. Moreover, the RL technique also has great potential to optimize a wireless system even without accurate or complete observation of the system state, which might happen in practical implementations. Particularly, the optimization of WiFi systems with implementation constraints would be investigated in this paper.

There have been a significant amount of works optimizing the throughput, delay or age-of-information (AoI) performance of wireless networks via the method of RL.\@
Most of these works assumed full knowledge of the system state in algorithm design. They could be applied to the systems, where the global system state could be collected at a centralized controller in time. On the other hand, RL was also utilized to optimize the performance of wireless systems with distributive transmission scheduling, e.g., wireless fidelity (WiFi) systems. For instance, an adaptive channel contention mechanism was proposed for WiFi systems in~\cite{kumar2021adaptive}, where a local RL agent was deployed at each user equipment (UE). The local agents adjusted the minimum contention window (MCW) size according to the global statistics of successful channel contention such that the transmission fairness among the agents can be ensured.
In order to resolve the collision issue of distributive channel access, deep RL algorithms were proposed in~\cite{ali2018deep} to determine the timing of doubling the contention window based on the estimated collision probability.
In addition to the adaptive channel contention, a double deep Q-network (DDQN)~\cite{van2016deep} based rate adaptation algorithm was proposed in~\cite{chen2021experience} to improve network throughput, where the agent inferred the optimal transmission rate based on the modulation and coding scheme (MCS) and frame loss rate. Most of the above literature assumed knowledge of the physical (PHY) layer and media access control (MAC) layer states. 
In fact, it might be challenging to obtain such knowledge in the scheduler design of a practical WiFi network. Moreover, the absence of knowledge on co-channel interference and the vendor-dependent implementation of WiFi adapters would also raise challenges.  In \cite{wang2024smuff}, the rate of WiFi direct transmission was optimized by managing the transmission and receiving buffer in user-space, instead of the MAC layer buffer. However, the scheduling of WiFi transmission with implementation constraints has not been addressed satisfactorily.

In this paper, we would like to shed some light on the RL-based scheduling design for practical WiFi systems suffering from unknown co-channel interference. Particularly, a framework, namely {\systemname}, is proposed for the scheduling of delay-sensitive tasks and file delivery tasks in the application layer. In {\systemname}, a controller collects the QoS observations of all the tasks periodically and determines the rate limitation and contention window size for all the transmitters such that the channel contention among them can be coordinated according to the overall QoS objective. Since there is no analytical model feasible for the above scheduling design, a novel Q-network is designed to make the scheduling decision. In order to accelerate the Q-learning, imitators are first proposed and trained to mimic the relationship between the scheduling action and QoS in different communication scenarios, respectively. Hence the Q-network can be trained with the imitators in an offline manner. It is shown by the experiments that the proposed framework can adapt to the variation of task number, interfering traffic, and link quality, and significantly outperforms the current EDCA mechanism defined in IEEE 802.11e.

\section{System Model\label{sec:system_model}}

\subsection{Deployment Scenario\label{sec:Deployment Scenario}}

The proposed {\systemname} system is deployed in a WiFi network with multiple connected access points (APs) and UEs working on the same channel. Denote the number of the devices, including the APs and UEs, in the WiFi network as $\devNum$, the set of these devices as $
\devSet =\{\dev_{\devi}|{\devi}=\iterDeviceSet\}$, and the communication link from the $i$-th device to the $j$-th one as the {$\linkIdx$}-th link ($\forall \dev_{\devi}, \dev_{\devj} \in \devSet$). The communication links can be from UE to AP, from AP to UE, or between UEs (i.e., WiFi Direct). We define $\linkset$ as the set of all communication links in the system and $\linkset_{\devi} $ as the set of communication links from the $\dev_{\devi}$-th device. As a remark, one UE could simultaneously maintain the communication links to the AP and other UEs, where the transmission of the infrastructure and WiFi Direct modes is separated in the time domain.

The data traffics raised by the applications of UEs in $\devSet$ are referred to as communication tasks in this paper. For example, the application projecting the screen of a mobile phone to a laptop via WiFi Direct will raise a {\it delay-sensitive task}, e.g., Miracast~\cite{WiFiDisplay2011}, where an application-layer packet (i.e., video frame) is generated and delivered periodically (the typical period is $16$ ms). Moreover, file sharing between two devices will raise a {\it file delivery task}. For the elaboration convenience, we define $\thruTS$ and $\rttTS$ as the universal sets of file delivery tasks and delay-sensitive tasks on the $\linkIdx$-th link, respectively. A task is in the inactive state if there is no packet arrival or buffered file at the transmitter.

Because of the transmission latency constraint, the delay-sensitive tasks should be scheduled with higher priority than the file delivery ones. Hence, all the transmitters access the channel via the enhanced distributed channel access (EDCA) mechanism defined in IEEE 802.11e. Particularly, four access category (AC) queues, namely voice (VI), video (VO), best effort (BE), and background (BK), are adopted at all the transmitters. The transmission priorities of the four AC queues are differentiated by values of arbitration inter-frame spacing (AIFS) and contention window (CW) size. As in the practical systems, the file delivery tasks are scheduled with the BE priority, and the delay-sensitive tasks are scheduled with the VI priority. The latter has smaller AIFS and CW size, leading to a larger successful probability in channel contention. As a remark, due to the distributive channel contention mechanism, it is infeasible to accurately control the packet transmission order among the devices of a WiFi network with commercial WiFi adapter. Instead, the packet transmission in the {\systemname} system is scheduled in a stochastic manner by adapting the CW sizes of AC queues in each device.

There are some other WiFi networks sharing the same channel in the coverage of the considered network. The traffic in these networks would degrade the QoS of the considered network, e.g., larger delivery latency and lower throughput. Denote the set of devices in the interfering networks as $\devSet_I$. The communications among the devices in $\devSet_I$, namely interfering traffic, cannot be scheduled by the {\systemname} system. Instead, the {\systemname} system is designed to deduce the interference level and adjust the transmission accordingly.

\subsection{Task Queuing Model}
For each file delivery task, all the information bits to be delivered are saved in an application-layer buffer, and a user datagram protocol (UDP) socket is established at the very beginning of transmission.
The data dispatch from the buffer to the UDP socket is controlled by a dispatcher.
The UDP socket encapsulates the received data from the dispatcher into UDP datagrams and forwards them to the driver of WiFi adapter.

As a remark, the new datagrams at the WiFi adapter may not be transmitted immediately.
In fact, each WiFi adapter maintains four MAC-layer AC queues associated with the four transmission priorities, respectively.
The arrival datagrams are saved in the corresponding queues and transmitted following unknown vendor's protocol.
The queuing status of the WiFi adapter is usually not accessible in the application-layer.
Thus, it is infeasible for the proposed system to know when the WiFi adapter completely delivers a datagram; it is, therefore, infeasible for the proposed system to precisely control the transmission of a UDP datagram or an application-layer packet. As a result, the scheduling of the proposed system is designed based on the average observable performance in the application layer.

Specifically, the transmission time is organized into a sequence of scheduling periods, each with a duration of $\slotlen$ seconds. $\slotlen$ is sufficiently large to accommodate a number of MAC protocol data unit transmissions. Due to the invisibility of adapter status, the QoS of a file delivery task is measured by its application-layer throughput in one scheduling period. Particularly, for the ${\m}$-th file delivery task of the $\linkIdx$-th link, its QoS in the $t$-th scheduling period {$\eleThru(t)$} is defined as the number of information bits transferred from the task buffer to the associated UDP socket.
The dispatcher is designed to adaptively limit the throughput of the file delivery task such that delay-sensitive tasks could have a larger chance to access the channel. 
Hence, let $\eleThruLimit(t)$ be the throughput limitation of the ${\m}$-th file delivery task of {$\linkIdx$}-th link in the {$t$}-th scheduling period, the dispatcher would make sure
\begin{align}
    \label{eq:throughput_limitation}
    \eleThru(t) \leq \eleThruLimit(t).
\end{align}

For each delay-sensitive task, a task queue and UDP socket are established at the very beginning. The application-layer packets arrive at the task queue periodically with a fixed average data rate. The first packet in the queue is forwarded to the UDP socket for WiFi transmission as long as the socket is idle. Due to the lack of MAC-layer status, the measurement of the transmission latency of a packet could hardly be accurate. Hence, we use the round-trip time (RTT) as the QoS measurement of delay-sensitive tasks. Particularly, for each delay-sensitive task, an acknowledgment will be sent back from the receiver to the transmitter when an application-layer packet is completely received. Hence, the transmitter can calculate the RTTs of all packet transmissions. For the {${\m}$}-th delay-sensitive task of the {$\linkIdx$}-th link ($\forall \iterLink, \iterTaskRTT, $), its QoS in the $t$-th scheduling period {$\eleRTT(t)$} is defined as the average RTT of the packets transmitted in this scheduling period.

\subsection{Scheduling Model}

Denote the CW sizes of the VI and BE priorities of the ${\devi}$-th device in $t$-th scheduling period as  $w_{\devi}^{\VI}(t)$ and $w_{\devi}^{\BE}(t)$ respectively, we focus on the joint scheduling of these channel contention parameters and the dispatchers' throughput limitation $\{\eleThruLimit(t)|\forall \iterLink,\iterTaskThru\}$ in each scheduling period.

Particularly, each transmitter collects the QoS observations of its tasks in the end of each scheduling period and delivers them to a centralized controller, which can be implemented in an AP or other device, for making scheduling decision. Not all the tasks in the universal task sets are in the active state. The average RTTs and throughputs of inactive delay-sensitive and file delivery tasks are denoted by a sufficiently large value and $0$, respectively. Hence, the aggregation of QoS observations received at the controller in the end of the $t$-th scheduling period can be represented as
\begin{align}
    \begin{split}
        \observation_{t} \triangleq & \left\{\eleThru(t) |\forall \iterLink,\iterTaskThru \right\} \\ &\cup  \left\{\eleRTT(t)| \forall \iterLink,\iterTaskRTT \right\}.
    \end{split}
\end{align}

Due to the time-varying traffic of the interfering devices, the scheduling parameters, including the file throughput limitations and CW sizes, are adapted at the centralized controller in each schedule according to the system's scheduling parameters and QoS observations in the past $\stateSize$ scheduling periods. Specifically, the aggregation of scheduling parameters in the $t$-th scheduling period ($\forall t$) is represented as
\begin{align}
    \label{eq:action}
    \begin{split}
        \action_t \triangleq & \left\{ \eleThruLimit(t) | \forall \iterLink,\iterTaskThru \right\} \\ &\cup  \left\{ w_{\devi}^{\VI}(t), w_{\devi}^{\BE}(t) | i = \iterDeviceSet \right\}.
    \end{split}
\end{align}
Thus, in the very beginning of the $t$-th scheduling period, $\action_t$  ($\forall t$) is determined based on past scheduling parameters and QoS observations 
$\{ (\observation_{t-\stateSize},\action_{t-\stateSize}),(\observation_{t-\stateSize+1},\action_{t-\stateSize+1}),\ldots,(\observation_{t-1},\action_{t-1})\}$.

\section{Problem Formulation}\label{sec: Problem Formulation}
The proposed {\systemname} system should successively make scheduling decisions for each scheduling period. Hence, it could be formulated as a Markov decision process (MDP). 

\begin{Definition}[System State] In the $t$-th scheduling period ($\forall t$), the system state is defined as the aggregation of the QoS observations and scheduling parameters of the past $\stateSize$ scheduling periods. Thus, $\state_t \triangleq \left\{
        (\observation_{t-\stateSize},\action_{t-\stateSize}),\ldots,(\observation_{t-1},\action_{t-1})
        \right\}$.
\end{Definition}

\begin{Definition}[Scheduling Action and Policy] Denote $\action_t$ in (\ref{eq:action}) as the action in the $t$-th scheduling period,
    $\action_t^i \triangleq 
    \left\{ 
        \thruLimit_\link^\m(t) | \forall \devj \in \linkset_\devi, \iterTaskThru 
    \right\} 
    \cup 
    \left\{ 
        w_{\devi}^{\VI}(t), w_{\devi}^{\BE}(t) 
    \right\}$, as the local action of the $i$-th device in the $t$-th scheduling period. The scheduling policy {$\policy$} is a mapping from {state space} to {action space} as
        $\policy( \state_t ) = \action_t$.
\end{Definition}

Moreover, the system cost of the $t$-th scheduling period is defined as
\begin{align}
    \label{eq:cost function}
    \begin{split}
        \cost_t(\state_t, \action_t)
        \triangleq
         & \sum_{\iterLink} \sum_{\iterTaskRTT}
        \mathds{1}(\eleRTT(t) > \eleRTTLimit)                          \\
         & - \weight \sum_{\iterLink} \sum_{\iterTaskThru} \eleThru(t),
    \end{split}
\end{align}
where $\weight$ is a weight, $\eleRTTLimit$ is the maximum tolerable RTT of the ${\m}$-th delay-sensitive task on the $\linkIdx$-th link. The indicator function $\mathds{1}(\mathcal{E})$ is $1$ if the event $\mathcal{E}$ is true, and $0$ otherwise.
Then, the overall system cost is defined as the average discounted summation of system costs for all the scheduling periods, i.e.,
\begin{equation}
    \label{eq:average_cost}
    \averageCost(\policy) = \lim_{T \rightarrow \infty} \mathbb{E}
    \bigg[
        \sum_{t = 1}^T \gamma^{t-1} \cost_t(\state_t, \policy(\state_t))
        \bigg].
\end{equation}
For the elaboration convenience, it is assumed that the system has run for at least $N$ scheduling periods before the first scheduling period, such that there are sufficient QoS observations in the system state. As a result, the controller design of the {\systemname} system can be formulated as
\begin{equation}
    \text{\bf Problem 1:} \ \min_{\policy} \  \averageCost(\policy).
\end{equation}
The Bellman's equations for the above MDP is given by
\begin{align}
    \label{eq:bellman}
    Q(\state_t, \action_t) = \mathbb{E}_{\state_{t+1}} \bigg[ \cost_t(\state_t, \action_t) + \gamma \min_{\action'} Q(\state_{t+1}, \action')  \bigg],
\end{align}
where $Q(\state_t, \action_t)$ is the Q-function with system state $\state_t$ and action $\action_t$. Moreover, the optimal scheduling is given by
\begin{equation}
    \policy^*(\state)  = \argmin\limits_{\action} Q(\state, \action).
\end{equation}

Given the system state, it is still difficult to accurately predict the relation between the scheduling action and task QoS in the current scheduling period. This is because of the unknown interfering traffic and random channel contention. It is therefore impossible to solve the above Bellman's equations without any trial on the network performance. We shall rely on the RL method to track the above unknown knowledge with the assistance of a preliminary observation dataset $\trainingDataSet$.

Particularly, before the optimization, the dataset $\trainingDataSet$ is collected from $\dataSetSize$ scheduling periods experiencing heterogeneous interfering traffic and link quality. Each of the scheduling periods (say the $\dataSetTau$-th one) is divided into two phases. In the first phase, a fixed testing scheduling action $\classifyAction$ is applied, and corresponding QoS observation $\classifyObs_\dataSetTau$ is obtained; in the second phase, a random scheduling action $\trainAction_{\dataSetTau}$ is applied, and another QoS observation $\trainObs_{\dataSetTau}$ is obtained. Hence,  the dataset $\trainingDataSet$  can be expressed as $\trainingDataSet \triangleq
    \left\{
    (\classifyObs_\dataSetTau, \classifyAction, \trainObs_{\dataSetTau}, \trainAction_{\dataSetTau})
    | \dataSetTau = \iterDataSet
    \right\}$.

\section{Q-Network for Online Scheduling\label{sec:Q-network for Online Scheduling}}

In this section, a novel Q-network design is proposed to approximate the Q-function.
In order to accelerate the convergence of training and improve the scheduling performance, all the possible system performance of one scheduling period is divided into $K$ regions, and the inputs of the Q-network include not only the system state but also the performance region indices of the past $\stateSize$ scheduling periods.

Hence, the utilization of the proposed Q-network in the transmission scheduling can be divided into two stages. In the first stage, namely the offline stage, the performance regions are trained via the preliminary observation dataset $\trainingDataSet$, and the Q-network is then trained via $\trainingDataSet$ in all the performance regions respectively. In the second stage, namely the online stage, the Q-network is applied to the transmission scheduling and fine-trained according to the online QoS observations.

In this section, the performance region quantization is introduced first, followed by the structure of the Q-network. The hybrid offline and online training of Q-network is elaborated in Section \ref{sec:Hybrid Q-Learning}.

\subsection{Performance Region Quantization\label{sec:performance Region Quantization}}
The QoS observations with the testing scheduling action $\classifyAction$ are first extracted from the preliminary observation dataset $\trainingDataSet$ as $\classifyDataSet \triangleq
   \left\{
   (\classifyObs_\dataSetTau, \classifyAction) | \dataSetTau = \iterDataSet
   \right\}$.
The $\classifyNum$-means classification method~\cite{macqueen1967some} is then adopted to classify the QoS observations in $\classifyDataSet$ into $\classifyNum$ clusters. Denote the mean and variance of the observed throughputs (for the file delivery tasks) in $\classifyDataSet$ as $\avgThru$ and $\varThru^2$ respectively, the mean and variance of the RTTs (for the delay-sensitive tasks) as {$\avgRTT$ and $\varRTT^2$} respectively. The performance region quantization can be achieved by finding the $\classifyNum$ cluster centers of the QoS observations in $\classifyDataSet$ as follows:
\begin{equation}
   \{\iterCCOpt\}  =
   \argmin
   \limits_{\iterCC} \
   \sum_{\classifyIdx=1}^\classifyNum
   \sum_{\dataSetTau=1}^\dataSetSize
   \Vert \vectorization(\classifyObs_\dataSetTau) - \classifyCenter_\classifyIdx \Vert^2,
\end{equation}
where $\vectorization(\classifyObs_\dataSetTau)$ denotes the vectorization of the normalized QoS observations in $\classifyObs_\dataSetTau$. Particularly,
   $\vectorization(\classifyObs_\dataSetTau) \triangleq
   \left(
   \mathbf{\thru}_{\dataSetTau}^\classifyFlag, \mathbf{\rtt}_{\dataSetTau}^\classifyFlag
   \right)$,
where the row vector $\mathbf{\thru}_{\dataSetTau}^\classifyFlag$ vectorizes the normalized throughputs of all file delivery tasks in $\classifyObs_\dataSetTau$,
\begin{equation*}
   \left\{
      \frac{ \eleThruClassify(\dataSetTau) - \avgThru }{ \varThru }
      \bigg|
      \forall \iterLink, \iterTaskThru, \eleThruClassify(\dataSetTau) \in \classifyObs_\dataSetTau
   \right\},
\end{equation*}
and the row vector $\mathbf{\rtt}_{\dataSetTau}^\classifyFlag$ vectorizes the normalized RTTs of all the delay-sensitive tasks in $\classifyObs_\dataSetTau$, 
\begin{equation*}
   \left\{
      \frac{ \eleRTTClassify(\dataSetTau) - \avgRTT }{ \varRTT }
      \bigg| 
      \forall \iterLink, \iterTaskRTT, \eleRTTClassify(\dataSetTau) \in \classifyObs_\dataSetTau
   \right\}.
\end{equation*}

With $\{\iterCCOpt\}$, the performance region index of a scheduling period can be determined according to 
\begin{equation}
   \label{eq:raw performance region index}
   \rawPerRegion =
   \argmin\limits_{\classifyIdx} \
   \Vert \vectorization(\widehat{\observation}) - \classifyCenter_\classifyIdx^* \Vert^2,
\end{equation}
where $\widehat{\observation}$ is the aggregation of QoS observations with the testing scheduling action $\classifyAction$ in the scheduling period.

\begin{Remark}
Note that the QoS observations of the testing scheduling action $\classifyAction$ should be collected to determine the performance region index of one scheduling period. In the online stage, one short period can be reversed in each scheduling period to apply the testing scheduling action $\classifyAction$. 
\end{Remark}

\subsection{Q-Network Structure\label{sec:Q-network}}

The input of the proposed Q-network is the extended system state of the current scheduling period, which is defined below:
\begin{Definition}[Extended System State]
   In the $t$-th scheduling period ($\forall t$) of either offline or online training, the extended system state consists of
   $
   \extendState_t \triangleq 
   \left\{
      (\rawPerRegion_{t-\stateSize}, \observation_{t-\stateSize},\action_{t-\stateSize})
      ,\ldots,
      (\rawPerRegion_{t-1}, \observation_{t-1}, \action_{t-1})
   \right\}
   $, where $\rawPerRegion_{t-i}$ ($i=1,2,...,N$) is the performance region index.
\end{Definition}

The first part of the Q-network is a multi-head attention layer\cite{vaswani2017attention},  which is trained to refine the performance region indices in the extended system state. The refined extended system state is then used as the input of the following three fully connected layers with $256$ nodes and ReLU activation function sequentially.

In order to address the issue of huge action space, we adopt the following linear approximation structure on the Q-function in the output of the Q-network:
\begin{equation}
   Q(\extendState, \action)
   \approx
   \sum_{{\devi} \in \devSet} Q^{\devi}(\extendState, \action^{\devi}),
\end{equation}
where $Q^{\devi}(\extendState, \action^{\devi})$ is referred to as the local Q-function of the ${\devi}$-th device. Hence, the Q-network output consists of $\devNum$ action clusters for $\devNum$ devices, respectively. Each action cluster provides the values of the corresponding local Q-function for all possible local actions. As a result, the optimized local action of the $i$-th device ($\forall i$) in the $t$-th scheduling period of either offline or online training can be obtained by minimizing the local Q-function, i.e.,
\begin{equation}
   \action^{\devi}_t = \argmin\limits_{\action^{\devi}} Q^{\devi}(\extendState_t, \action^{\devi}).
\end{equation}

\section{Hybrid Q-Learning\label{sec:Hybrid Q-Learning}}
\begin{figure*}[!t]
   \normalsize
   \begin{equation}
      \label{eq: imitator loss}
      \imitatorLoss(\inetParam_\classifyIdx) = \frac{1}{|\trainObs_{\dataSetTau}|} \left[
         \alpha
         \sum_{\iterLink} \sum_{\iterTaskThru}
         {\left(\elePreThru(\action; \inetParam_\classifyIdx) - \eleThru(\dataSetTau)\right)}^2
         +
         \sum_{\iterLink} \sum_{\iterTaskRTTAlt}
         {\left(
            \elePreRTT(\action; \inetParam_\classifyIdx)
            -
            \min \left\{
            \eleRTTAlt(\dataSetTau), \beta \rttLimit^{\n}_{\link}
            \right\}
         \right)}^2
         \right].
   \end{equation}
   \begin{equation}
      \label{eq:loss}
      \ctlLoss(\qnetParam_t) = \mathbb{E} \left[
         {\left( \cost_t\left(\state_t,\action_t\right)
         + \gamma \sum_{{\devi} \in \devSet}
         \min_{{\action^{{\devi}}}^'} Q^{\devi} \left( \extendState_{t+1}, \action^{{\devi}^'}; \qtnetParam_t \right)
         - \sum_{{\devi} \in \devSet}
         Q^{\devi} \left(
         \extendState_t, \action^{\devi}_t; \qnetParam_t
         \right)
         \right)}^2
         \right].
   \end{equation}
   \hrulefill{}
   \vspace{-15pt}
\end{figure*}

The Q-network is first trained in the offline stage based on the dataset $\trainingDataSet$, then tuned in the online stage. 
\subsection{Offline Imitation Learning and Q-Network Training}
To facilitate the offline training, the performance indices are calculated for all the scheduling periods in $\trainingDataSet$ according to  (\ref{eq:raw performance region index}). Denote the performance index of the $\dataSetTau$-th scheduling period in $\trainingDataSet$ as $\rawPerRegion_\dataSetTau^\trainFlag$, the preliminary dataset $\trainingDataSet$ can be rewritten as
\begin{equation}
   \trainingDataSetwithRawPerInd  \triangleq
   \left\{
   (\rawPerRegion_\dataSetTau^\trainFlag, \observation_{\dataSetTau}^\trainFlag, \action^s_{\dataSetTau})
   |
   \dataSetTau = \iterDataSet
   \right\}
\end{equation}
for notation convenience. Moreover, dataset $\trainingDataSetwithRawPerInd$ can be further divided into $\classifyNum$ subsets as
\begin{equation}
   \trainingDataSetwithRawPerInd_\classifyIdx
   \triangleq \left\{
   (\classifyIdx, \observation_{\dataSetTau}^\trainFlag, \action^\trainFlag_{\dataSetTau})
   |
   \forall \rawPerRegion_\dataSetTau^\trainFlag = \classifyIdx
   \right\}
   \subset
   \trainingDataSetwithRawPerInd, \classifyIdx = 1,\ldots, \classifyNum.
\end{equation}

Notice that the subsets $\trainingDataSetwithRawPerInd_\classifyIdx$ ($\classifyIdx=\iterCCIdx$) may not be sufficiently large for the training of the Q-network in all the performance regions, the imitation learning method is introduced. Particularly, we first train $\classifyNum$ DNN networks (namely imitators), each of which consists of $10$ fully connected layers and $256$ nodes per layer, to imitate the relation between the scheduling actions and QoS observations in the $\classifyNum$ performance regions, respectively. Denote the imitators as $f(\action; \inetParam_\classifyIdx), \classifyIdx=\iterCCIdx$, where $\action$ is the input action, and $\inetParam_\classifyIdx$ represents network parameters. The output of imitator $f(\action; \inetParam_\classifyIdx)$ is trained to approximate the QoS observations of the system in the $\classifyIdx$-th performance region with input action $\action$. Then, the Q-network can be trained via the $\classifyNum$ imitators.

{\bf Imitator training:} The $k$-th imitator ($k=1,2,...,K$) is trained by $\trainingDataSetwithRawPerInd_\classifyIdx$. Let $\elePreThru(\action; \inetParam_\classifyIdx)$ and $\elePreRTT(\action; \inetParam_\classifyIdx)$ be the throughput and RTT of the ${\m}$-th file delivery task and ${\n}$-th delay-sensitive task of the $\linkIdx$-th link in the output of the $\classifyIdx$-th imitator with input action $\action$.
The loss function $\imitatorLoss$ is defined as (\ref{eq: imitator loss}), where $\thru^{\m}_{\link}(\dataSetTau),\rtt^{\n}_{\link}(\dataSetTau) \in \observation_{\dataSetTau}^s$,  $\alpha$ and $\beta$ are both weights, and the minimization is to limit the range of RTTs.

{\bf Offline Q-network training:}  Based on the imitators, the Q-network can be trained in each performance region respectively. Particularly, in the $t$-th scheduling period of offline training with the $k$-th imitator ($\forall t,k$), providing the scheduling action, the outputs of the imitator are treated as the QoS observations in the $\classifyIdx$-th performance region, which is then used to update the extended system state of the $(t+1)$-th scheduling period in the input of the Q-network. The Q-network is also updated in the above iterative procedure according to the Q-learning method~\cite{mnih2015human}. The loss function 
$\ctlLoss$ is defined in (\ref{eq:loss}), where $Q(\cdot, \cdot; \qnetParam_{t})$ represents the Q-network parameters in the $t$-th scheduling period, and $\qtnetParam_t$ denotes the parameter of target network as in \cite{mnih2015human}. 

In order to efficiently explore the action space, an upper confidence bound (UCB) based exploration policy is introduced to determine the scheduling action in the offline training of Q-network. Taking the $t$-th scheduling period with the $\classifyIdx$-th imitator as the example, we first define the UCB of the action $\action^{\devi}$ of ${\devi}$-th device as
\begin{equation}
   \label{eq:UCB}
   UCB_{t}(\classifyIdx, \action^{\devi}) =
   Q_{t}^{\devi}(\extendState_t, \action^{\devi}; \qnetParam_{t})
   +
   \sqrt{\frac{4\eta\ln t}{T_{t}( \classifyIdx , \action^{\devi})}},
\end{equation}
where $T_t(\classifyIdx,\action^{\devi})$ counts the number of times the action $\action^{\devi}$ is taken up to the $t$-th scheduling period. The hyper-parameter $\eta$ is used to balance the exploration and exploitation.
As a result, the scheduling action is determined as follows:
\begin{equation}
   \label{eq:UCB epolicy}
   \action_t^{\devi} = \begin{cases}
      \argmin UCB_{t}(\classifyIdx, \action^{\devi})         & \text{w.p. } 1 - \epsilon_t, \\
      \action^{\devi} \sim \uniformDis(\actionSpace^{\devi}) & \text{w.p. } \epsilon_t,
   \end{cases}
\end{equation}
where $\uniformDis(\actionSpace^{\devi})$ is the uniform distribution over action space $\actionSpace^{\devi}$ of ${\devi}$-th device and exploration rate $\epsilon_t$ should satisfy the limit condition $\lim_{t \to \infty} \epsilon_t = 0$.

\subsection{Online Q-network Training} 
The online Q-network training with the same loss function as in (\ref{eq:loss}) could be applied to further improve the performance of the proposed {\systemname} system. Particularly, in the $t$-th scheduling period of the online stage, the scheduling action of the ${\devi}$-th device, denoted as $\action_t^i$, is determined by the $\epsilon$-greedy policy as follows:
\begin{equation}
   \label{eq:online epolicy}
   \action_t^{\devi} = \begin{cases}
      \argmin Q^{\devi}(\extendState_t, \action^{\devi}; \netParam_{t}) & \text{w.p. } 1 - \epsilon_t, \\
      \action^{\devi} \sim \uniformDis(\actionSpace^{\devi})            & \text{w.p. } \epsilon_t,
   \end{cases}
\end{equation}
where $\epsilon_t$ and $\uniformDis(\actionSpace^{\devi})$ are defined in (\ref{eq:UCB epolicy}).

\section{Experiments}
\label{sec:testbed} %

\begin{figure}[!t]
    \centering
    \includegraphics[width=0.8\linewidth]{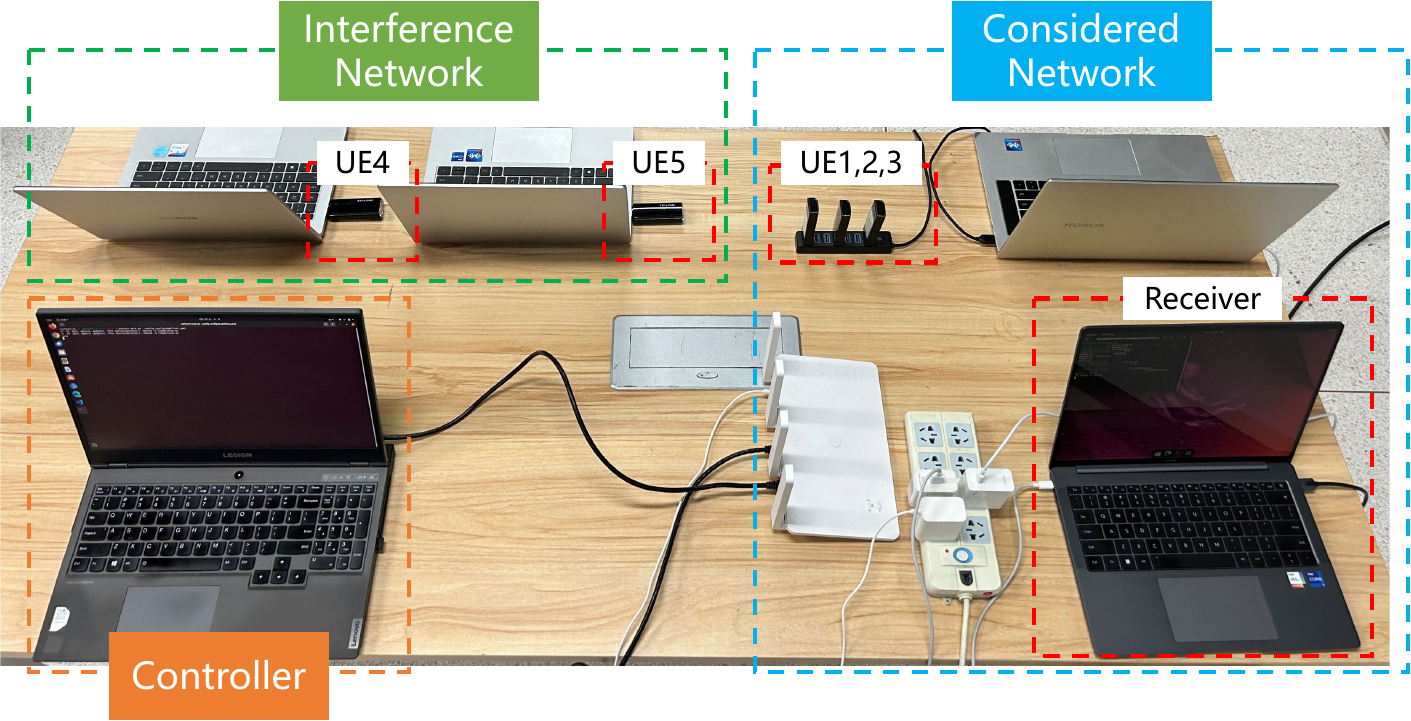}
    \caption{Illustration of testbed.}
    \label{fig:Testbed Setup} %
\end{figure}

As illustrated in Fig. \ref{fig:Testbed Setup} The proposed {\systemname} system is implemented in a WiFi network with one HONOR XD30 AP and $3$ UEs each equipped with a TP-Link TL-WDN6200 USB WiFi adapter in the experiment\footnote{The demo video is available in \href{http://lasso.eee.sustech.edu.cn/reinwifi/}{http://lasso.eee.sustech.edu.cn/reinwifi/};
The source code is available in \href{https://github.com/QianrenLi/ReinWiFi}{https://github.com/QianrenLi/ReinWiFi}.}. 
Denote the AP as $\dev_0$ and the three UEs as $\dev_1, \dev_2, \dev_3$, respectively. 
The network is working on the $5$G WiFi band following the IEEE 802.11ac specification.
The real-time controller is implemented in a laptop with Intel Core i7-8750H CPU and Ubuntu 20.04 operating system. %
An Ethernet connection with a $1$ Gbps data rate is employed to facilitate communication between the controller and the AP.\@ 
Moreover, we implement a Linux module to adapt the CW sizes of adapters in real-time from user space. 
The controller can collect the QoS observations from
UEs and notify the scheduling actions via WiFi, such that the UEs' transmission scheduling can be adjusted accordingly. 

Both file delivery tasks and delay-sensitive tasks are tested in the experiment. The former tasks with a sufficient backlog are transmitted with the BE priority. The latter tasks, consisting of two types, are delivered with the VI priority. The data rates of type I and II delay-sensitive tasks are $\arrival_1 = 50$Mbps and $\arrival_2 = 25$Mbps, respectively. The packet arrival intervals of the two types are both  $16$ ms. Moreover, the maximum tolerable RTTs are $16$ms and $28$ms, respectively. The universal set of communication tasks tested in the experiment includes a delay-sensitive task with arrival data rate $\arrival_1$ (Task 1) and a file delivery task (Task 2) on the $(\firExpLink)$-th link; a delay-sensitive task with arrival data rate $\arrival_2$ (Task 3) on the $(\secExpLink)$-th link; a delay-sensitive task with arrival data rate $\arrival_2$ (Task 4) on the $(\thiExpLink)$-th link. The quality of the $(\firExpLink)$-th, $(\secExpLink)$-th, and $(\thiExpLink)$-th links depend on their distances and the propagation environment, which could be changed in the experiment. 

In the experiment, the scheduling period duration is 1 second, the CW size takes values from $\{2^{\counter} - 1 \mid \counter = 1,2,\ldots,10\}$, and throughput limitation takes values from  $\{ \frac{\counter}{20} \thru_{\link}^{{\m},\max} \mid \counter =0,1,\ldots,20\}$, where $\thru_{\link}^{{\m},\max}$ = 600 Mbps. Moreover, in addition to the background interference, the interfering traffic between two interference UEs, namely $\dev_4$ and $\dev_5$, is generated with a random data rate and BE priority in the same channel.

The preliminary observation dataset $\trainingDataSet$ is collected from the following three different traffic patterns (TPs): (1) Tasks 1 and 2 are activated; (2) Tasks 1, 2, and 3 are activated; and (3) Tasks 1, 2, 3 and 4 are activated. In all the TPs, the communication distances of the links are altered to exploit the diversity of link rates. In the collection of $\trainingDataSet$, the testing scheduling action $\action^\classifyFlag$ is first applied in the first half of the scheduling period, where the CW size and throughput limitations are $7$ and $300$ Mbps respectively. Then, a randomized action is applied in the second half. QoS observations of both actions are collected in each scheduling period.

Based on dataset $\mathscr{S}^s$, the performance of the three TPs are quantized into 3, 6, and 6 regions, respectively. Then, $15$ QoS imitators are trained according to Section~\ref{sec:Hybrid Q-Learning} with $\alpha=1$, $\beta=3$. Given the trained QoS imitators, the Q-network is further trained as elaborated in Section~\ref{sec:Hybrid Q-Learning} with $\weight=1/{r}^{m,\max}_{\link}$.

To demonstrate the performance gain, the proposed framework is compared with two baselines. The first baseline, namely {\it Standard EDCA}, relies on the conventional 802.11 EDCA protocol. The second baseline, namely {\it Rate Control Only}, adapts the throughput limitation of file delivery tasks via the proposed framework with the CW sizes following the 802.11 EDCA protocol.\@
The performance evaluation and comparison are conducted in $11$ distinct test scenarios listed in Table~\ref{tab:Test Scenario}, where only the first $5$ scenarios have been measured in the preliminary observation dataset $\trainingDataSet$.

\begin{table}
    \vspace{3pt}
    \centering
    \begin{tabular}{c|c|c||c|c|c} %
    \specialrule{.1em}{.05em}{.05em} 
    Scenario  
    & 
    TP                        
    & 
    \begin{tabular}[c]{@{}c@{}}Link Rate\\(Mbps)\end{tabular}  
    &
    Scenario  
    & 
    TP                        
    & 
    \begin{tabular}[c]{@{}c@{}}Link Rate\\(Mbps)\end{tabular}  
    \\ 
    \hline %
    1   & 1    &  563, 499, 572  & 7   & 3    &  563, 424, 572       \\ \hline         %
    2   & 2    &  563, 499, 572  & 8   & 2    &  563, 400, 346       \\ \hline         %
    3   & 3    &  563, 499, 572  & 9   & 3    &  563, 400, 346       \\ \hline         %
    4   & 2    &  563, 370, 572  & 10  & 2    &  459, 499, 572       \\ \hline         %
    5   & 3    &  563, 370, 572  & 11  & 3    &  459, 499, 572       \\ \hline         %
    6   & 3    &  563, 499, 476  &&&       \\ \hline         %
    \end{tabular}
    \caption{Table of test scenarios, where the link rate refers to the maximum data rates of the $(\dev_1,\dev_0)$-th, $(\dev_2,\dev_0)$-th, and $(\dev_3,\dev_0)$-th links.\label{tab:Test Scenario}}
\end{table}

\begin{figure}
    \centering
    \includegraphics[clip,trim=0cm 0cm -0.3cm 0cm, width=\linewidth]{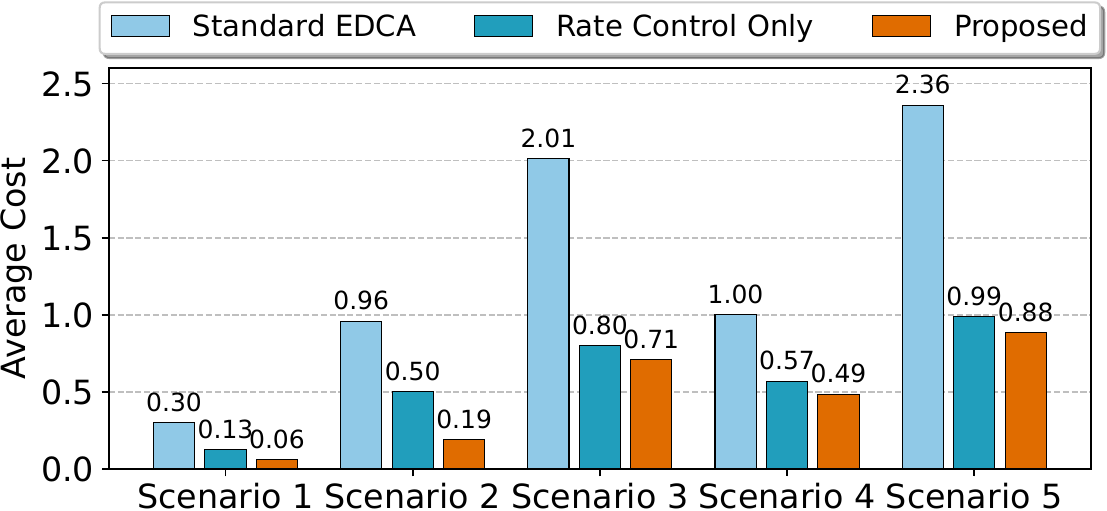}
    \caption{Performance comparison in scenarios $1 \sim 5$.\label{fig:Average cost comparison of proposed algorithm in Scenario 1, 2, 3, 4 and 5.}}
\end{figure}

The performance comparison of the proposed framework and the two baselines in the first $5$ test scenarios is illustrated in Fig.~\ref{fig:Average cost comparison of proposed algorithm in Scenario 1, 2, 3, 4 and 5.}, where the online training is not applied in the proposed framework and the Baseline 2. It can be observed that the proposed Q-network offline trained via imitators significantly outperforms the conventional EDCA mechanism. Moreover, the performance gain of the Baseline 2 over Baseline 1 demonstrates the necessity of the throughput limitation, which has never been investigated in the existing literature. 

The performance comparison in the test scenarios $6$ to $11$ is illustrated in Fig.~\ref{fig:Test-changing interference-cost for test scenario 6, 7 and 8.}. Since these test scenarios are not measured in the preliminary observation dataset $\trainingDataSet$, the performance gain of the proposed scheme over the Baseline 1 demonstrates the good generalization capability of the proposed Q-network. It can also be observed that the online training could further improve the scheduling performance of the Q-network, which has already been trained in the offline stage.

\begin{figure}[!t]
    \centering
    \includegraphics[width=\linewidth]{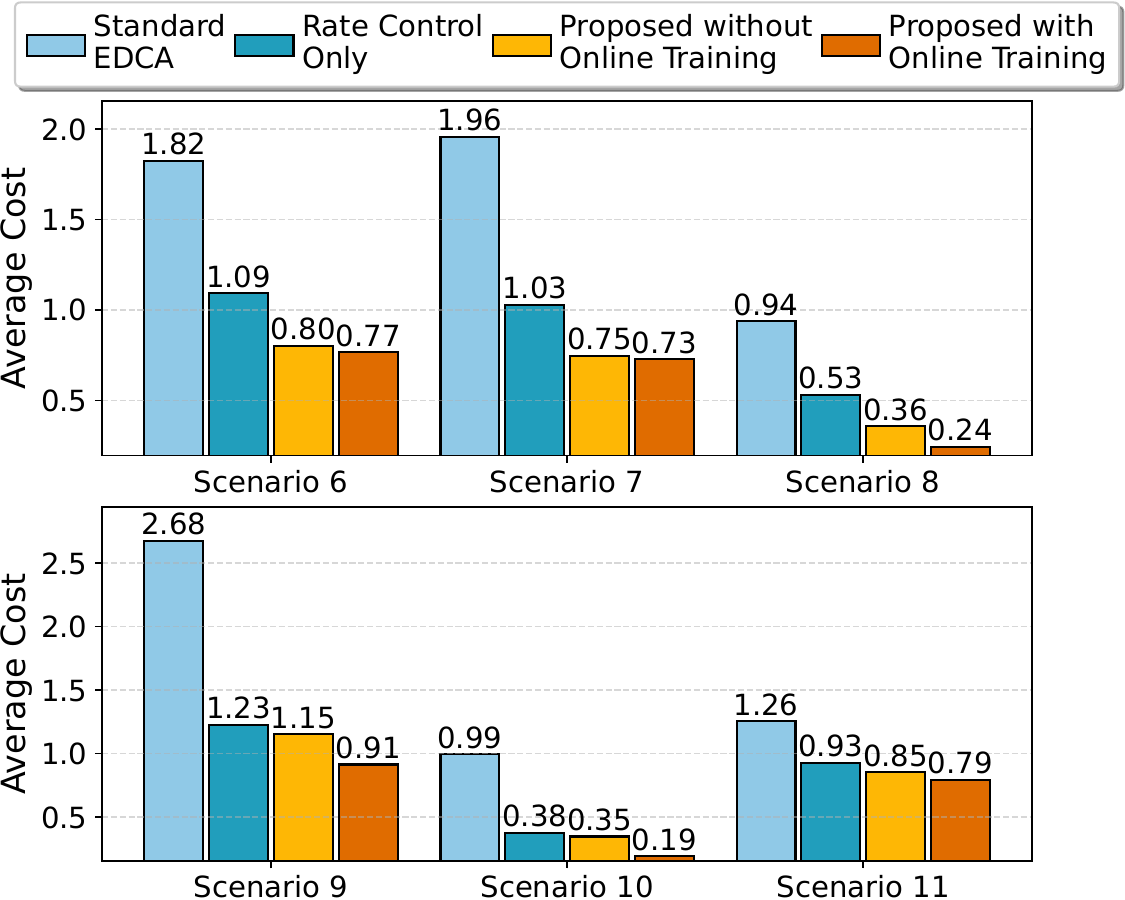} %
    \caption{Performance comparison in scenarios $6 \sim 11$.\label{fig:Test-changing interference-cost for test scenario 6, 7 and 8.}}
\end{figure}

\section{Conclusion\label{sec:Conclusion}}

In this paper, a reinforcement-learning-based framework, namely {\systemname}, is proposed for the application-layer QoS optimization of WiFi networks. Due to the absence of PHY-layer and MAC-layer status, the historical scheduling parameters and QoS observations are considered as the system state in the determination of the current scheduling parameters. Because of the unknown interference and vendor-dependent implementations, a novel Q-network is proposed to track the relation between the system state, scheduling parameter, and the overall QoS. Moreover, an imitation learning method is introduced to improve the training efficiency. It is demonstrated via the testbed that the proposed framework, with the dynamic adaptation of CW size and throughput limitation, significantly outperforms the convention EDCA mechanism.

\bibliographystyle{IEEEtran}
\bibliography{IEEEabrv,reference, abbrv}
\end{document}